\documentclass{revtex4}
\usepackage{amsmath,amsthm}
\usepackage{rotating}
\usepackage{graphicx}

\usepackage{amsfonts}
\newtheorem{theorem}{Theorem}[section]

\newtheorem{proposition}[theorem]{Proposition}

\theoremstyle{remark}

\theoremstyle{definition}

\theoremstyle{example}
\newtheorem{example}[theorem]{Example}

\theoremstyle{notation}

\newcommand{\bra}[1]{\langle#1|}
\newcommand{\ket}[1]{|#1\rangle}
\begin{document}

\title{ Markov chains with doubly stochastic transition matrices and application to a sequence of non-selective quantum measurements}            
\author{A. Vourdas}
\affiliation{Department of Computer Science,\\
University of Bradford, \\
Bradford BD7 1DP, United Kingdom\\a.vourdas@bradford.ac.uk}

\begin{abstract}
A time-dependent finite-state Markov chain that uses doubly stochastic transition matrices, is considered.
Entropic quantities that describe the randomness of the probability vectors, and also
the randomness of the discrete paths, are studied.
Universal convex polytopes are introduced which contain all
future probability vectors, and which are based on the Birkhoff-von Neumann expansion for doubly stochastic matrices.
They are universal in the sense that they depend only on the present probability vector, and are independent of the doubly stochastic transition matrices that describe time evolution in the future.
It is shown that as the discrete time increases these convex polytopes shrink,
and the minimum entropy of the probability vectors in them increases.
These ideas are applied to a sequence of non-selective measurements (with different projectors in each step) on a quantum system with $d$-dimensional Hilbert space.
The unitary time evolution in the intervals between the measurements, is taken into account. 
The non-selective measurements destroy stroboscopically the non-diagonal elements in the density matrix.
This `hermaphrodite' system is an interesting combination of a classical probabilistic system (immediately after the measurements) and a quantum system (in the intervals between the measurements).
Various examples are discussed. In the ergodic example, the system follows asymptotically all discrete  paths with the same probability.
In the example of rapidly repeated non-selective measurements, we get the well known quantum Zeno effect with `frozen discrete paths' (presented here as a biproduct of our general methodology 
based on Markov chains with doubly stochastic transition matrices).
\end{abstract}
\maketitle

\section{Introduction}

There are many processes with the Markov property in Science and Engineering  and also in Social Sciences, and there is a vast amount of literature that discusses these processes (e.g., \cite{M1,M2}).
Motivated by an application to quantum systems with non-selecitve measurements, we study  a time-dependent finite-state Markov chain  with doubly stochastic transition matrices.
Markov chains usually involve the more general row Markov transition matrices. 

The novel result in our case with doubly stochastic transition matrices, is the universal convex polytopes. They contain all
future probability vectors, and are based on the Birkhoff-von Neumann expansion for doubly stochastic matrices.
The universal convex polytopes depend only on the present probability vector, and are independent of the doubly stochastic matrices that describe time evolution in the future.
As the discrete time increases these convex polytopes shrink.

In this context, we use entropic quantities to quantify  the randomness of the probability vectors.
The Kullback-Leibler distance quantifies the distance of the probability vectors from the `uncertain' probability vector (in Eq.(\ref{5}) below).
We also introduce `discrete paths' and quantify their randomness with entropic quantities.
Ergodic Markov chains are an important special case, which is studied in this context.

These ideas are applied to a quantum system with $d$-dimensional Hilbert space (e.g.,\cite{V0,V1,V11}), on which we perform a sequence of non-selective measurements (with different projectors in each step).
Non-selective measurements have been used recently in quantum communications\cite{N,N0}, quantum control\cite{N1}, quantum Zeno effect \cite{Z1,Z2,Z3,Z4}, quantum thermodynamics\cite{N2,N3},  etc.

Between the non-selective measurements our system evolves with unitary transformations.
The non-selective measurements destroy stroboscopically the non-diagonal elements in the density matrix, and make the system a probabilistic mixture of states.
In each step we use different projectors, and therefore the non-diagonal elements refer to different bases.
There is a fundamental difference between a quantum system with superposition of states, and a classical probabilistic system with a probabilistic mixture of states.
The present system  belongs to the former category in the intervals between measurements, and to the latter category at the times of the measurements.
In this sense it is a `hermaphrodite' system that combines a classical probabilistic system with a quantum system.
This is a very different concept from a semiclassical system ($\hbar \rightarrow 0$).

Quantum Mechanics is not a Markovian theory (e.g., \cite{GI}).
This is related to the non-diagonal terms in the density matrix, which in turn are related to the superposition principle.
We note that there are various modifications of Markov models, called `quantum Markov models' (e.g., \cite{AC1,AC2,AC3,AC4,AC5}), which are suitable for the description of 
open quantum systems.
Our statement that quantum mechanics is not a Markovian theory refers to the `standard' (non-quantum) Markov models.
The difference between classical Markov models and quantum models in the context of probabilities related to social sciences, is discussed in \cite{SO}.

The quantum system  we study is forced by the non-selective measurements to be stroboscopically in a probabilistic mixture of states.
Between successive measurements the system evolves in a unitary quantum mechanical manner.
We show that the probability vectors associated to the measurements are described by a Markov chain that involves 
doubly stochastic transition matrices.

In section 2  we introduce briefly  doubly stochastic matrices and  permutation matrices.
In section 3 we discuss time-dependent finite-state Markov chains with doubly stochastic matrices.
Entropic quantities are used to describe the randomness of the probability vectors and of the discrete paths (in sections \ref{secQ1}, \ref{secQ2}, \ref{secQ3}). 
Ergodic Markov chains with doubly stochastic matrices, where the system follows asymptotically all paths with the same probability, are discussed in section \ref{sec89}.

An important aspect of our discussion on Markov chains, is the universal convex polytopes which contain all future probability vectors (in section \ref{sec450}).
They are universal in the sense that they depend only on the present probability vector, and they do not depend on the doubly stochastic transition matrices used in the Markov chain in the future.
Such polytopes are defined only for the special case of Markov chains with doubly stochastic transition matrices, and for this reason cannot be found in the literature on Markov chains.
We show that as the step number increases, the convex polytopes shrink and the minimum entropy of the probability vectors in them increases.

In section 4 we first show that  due to the non-diagonal elements in the density matrix, quantum Mechanics is not a Markovian theory (see also \cite{GI}).
We then consider a sequence of non-selective measurements together with the unitary evolution transformations between them.
We show that this can be described as a Markov chain that involves doubly stochastic matrices related to the unitary time evolution matrices. 
We also define universal convex polytopes that contain all future probability vectors associated with the measurements.

In section 5 we discuss various examples.
We consider a system which is initially in a position state, in a momentum state, and in the state $\rho=\frac{1}{d}{\bf 1}$.
We also discuss the homogeneous case, and the ergodic case.
We also discuss the special case of rapidly repeated non-selective measurements.
This leads to the well known quantum Zeno effect (`freezing of the paths'), which is  presented here with our methodology based on Markov chains with doubly stochastic transition matrices.
We conclude in section 6, with a discussion of our results.

\section{Doubly stochastic matrices}

Let ${\mathfrak D}$ be a $d\times d$ matrix with non-negative elements. It is a doubly stochastic matrix if 
\begin{eqnarray}\label{A7}
\sum _{a=1}^d{\mathfrak D}(a,b)=\sum _{b=1}^d{\mathfrak D}(a,b)=1;\;\;\;{\mathfrak D}(a,b)\ge 0.
\end{eqnarray}

The product of two doubly stochastic matrices is a  doubly stochastic matrix.
The inverse of a doubly stochastic matrix might not exist, or if it exists it might not be a doubly stochastic matrix.
The set of doubly stochastic matrices  is a  semigroup with respect to matrix multiplication.
Physically this means that processes that involve doubly stochastic matrices are irreversible.

The matrix ${\mathfrak U}$ with all elements
\begin{eqnarray}\label{UUU}
{\mathfrak U}(a,b)=\frac{1}{d};\;\;\;{\rm rank}({\mathfrak U})=1,
\end{eqnarray}
is doubly stochastic. 
For any doubly stochastic matrix ${\mathfrak D}$ we get 
\begin{eqnarray}
{\mathfrak U}{\mathfrak D}={\mathfrak D}{\mathfrak U}={\mathfrak U}.
\end{eqnarray}

Let ${\bf x}$ be a probability vector
\begin{eqnarray}
{\bf x}=(x_0,...,x_{d-1});\;\;\;x_i\ge 0;\;\;\;\sum _{i=0}^{d-1}x_i=1.
\end{eqnarray}
The product ${\bf x}{\mathfrak D}$ of a probability vector ${\bf x}$ (written as a row) times a doubly stochastic matrix ${\mathfrak D}$ (on the right), is also a probability vector 
(written as a row).
We note that the convention in the area of Markov processes is to multiply row vectors with matrices on the right.

A special case of a probability vector, is the most uncertain vector
\begin{eqnarray}\label{5}
{\bf u}=(u_0,...,u_{d-1});\;\;\;u_i=\frac{1}{d}.
\end{eqnarray}
For any doubly stochastic matrix ${\mathfrak D}$ we get
\begin{eqnarray}\label{6}
{\bf u}{\mathfrak D}={\bf u}.
\end{eqnarray}
Also for any probability vector ${\bf x}$ we get
\begin{eqnarray}
{\bf x}{\mathfrak U}={\bf u}.
\end{eqnarray}

A doubly stochastic matrix has the eigenvalue $e_0=1$, and the other eigenvalues $e_1,...,e_{d-1}$ have absolute value $|e_i|\le 1$.
The eigenvector corresponding to the eigenvalue $e_0=1$, is ${\bf u}^T$.
In particular, the matrix ${\mathfrak U}$  has the eigenvalues $e_0=1$ and $e_a=0$ for $a=1,...,d-1$. 
Therefore any doubly stochastic matrix can be written as
\begin{eqnarray}\label{8}
{\mathfrak D}={\bf u}^T{\bf u}+\sum _{i=1}^{d-1}e_i{\bf v_i}^T{\bf v_i}={\mathfrak U}+\sum _{i=1}^{d-1}e_i{\bf v_i}^T{\bf v_i};\;\;\;|e_i|\le 1,
\end{eqnarray}
where ${\bf v_i}^T$ are the eigenvectors corresponding to the eigenvalue $e_i$.

We define the entropy of a probability vector as
\begin{eqnarray}
E({\bf x})=-\sum _ax_a\log x_a;\;\;\;0\le E({\bf x})\le \log d
\end{eqnarray} 
$E({\bf x})=0$ for the most certain vector that has one component equal to one, and the other components equal to zero.
$E({\bf x})=\log d$ for the most uncertain vector ${\bf u}$.

In the set ${\cal P}$ of all probability vectors we define the `more certain' preorder\cite{MAJ,MAJ1}:
\begin{eqnarray}
{\bf x}\succ {\bf y}\;\leftrightarrow  E({\bf x})\le E({\bf y})
\end{eqnarray}
It is easily seen that
\begin{eqnarray}
&&{\bf x}\succ {\bf x}\nonumber\\
&&{\bf x}\succ {\bf y}\;{\rm and}\;{\bf y}\succ {\bf z}\;\rightarrow\;{\bf x}\succ {\bf z}.
\end{eqnarray}
Therefore $\succ$ is a preorder in ${\cal P}$.
We note that there are many quantities which are associated with this preorder (Schur concave functions) but in this paper we are interested only in the entropy.
It is known \cite{MAJ,MAJ1} that for any doubly stochastic matrix ${\mathfrak D}$,
\begin{eqnarray}\label{11}
{\bf x}\succ {\bf x}{\mathfrak D}.
\end{eqnarray}
Also ${\bf x}\succ {\bf u}$, for any ${\bf x}$.

\subsection{Permutation matrices: a special case of doubly stochastic matrices}
Let $\pi$ be a permutation of the integers $0,...,d-1$:
\begin{eqnarray}
(0,1,...,d-1)\;\overset{\pi} \rightarrow\;(\pi(0),\pi(1),...,\pi(d-1)).
\end{eqnarray}
The set of all permutations is the symmetric group ${\cal S}$\cite{SYM} which has $d!$ elements.
Multiplication in this group is the composition:
\begin{eqnarray}
(0,1,...,d-1)\;\overset{\pi _1} \rightarrow\;(\pi_1(0),\pi_1(1),...,\pi_1(d-1))\overset{\pi_2} \rightarrow\;(\pi_2 [\pi_1(0)],\pi_2[\pi_1(1)],...,\pi_2[\pi_1(d-1)]).
\end{eqnarray}
We will use the notation $(\pi_2 \circ \pi_1)(i)=\pi_2[\pi_1(i)]$.
The unity is
\begin{eqnarray}
(0,1,...,d-1)\;\overset{\bf 1} \rightarrow\;(0,1,...,d-1).
\end{eqnarray}

A permutation matrix $M_\pi$ is a $d\times d$ matrix with elements 
\begin{eqnarray}\label{PP}
M_\pi(a,b)=\delta(\pi(a),b);\;\;\;{\rm rank} (M_\pi)=d.
\end{eqnarray}
where $\delta$ is the Kronecker delta. 
Each row and each column have one element equal to $1$ and the other $d-1$ elements equal to $0$.
Permutation matrices are special cases of the doubly stochastic matrices.
It is easily seen that
\begin{eqnarray}
M_{\pi_2} M_{\pi_1}=M_{\pi_1\circ \pi_2};\;\;\;M_{\bf 1}={\bf 1};\;\;\;M_{\pi^{-1}}=[M_\pi]^{-1}=[M_\pi]^T.
\end{eqnarray}
The matrices $M_\pi$ with matrix multiplication form a representation of the symmetric group ${\cal S}$\cite{SYM}.

The Birkhoff-von Neumann theorem states that every doubly stochastic matrix  can be expanded in terms of the $d!$ permutation matrices with probabilities 
$\lambda(\pi)$ as coefficients:
\begin{eqnarray}
{\mathfrak D}=\sum _{\pi}\lambda(\pi)M_\pi;\;\;\;\sum _{\pi}\lambda(\pi)=1;\;\;\;\lambda(\pi)\ge 0.
\end{eqnarray}
This expansion is not unique. There is much work on the numerical calculation of the coefficients, which however is not relevant here.

If ${\mathfrak D}_1, {\mathfrak D}_2$ are doubly stochastic matrices, then the $p{\mathfrak D}_1+(1-p) {\mathfrak D}_2$ where $0\le p\le 1$ is a doubly stochastic matrix.
The set  of all $d\times d$ doubly stochastic matrices is a convex polytope with the $d!$ permutation matrices (in the symmetric group ${\cal S}$) as vertices.
It is known as Birkhoff polytope and is used in areas like Operational Research, Optimization, Enumerative Combinatorics, etc.

General references on convex polytopes are \cite{POLY1,POLY2}, and in a quantum context \cite{POLY3}. 
Numerical work to find the vertices of a convex polytope is discussed in \cite{VER}.

\section{Markov chains with doubly stochastic matrices}\label{sec56}

In a finite-state Markov chain the system is in one of $d$ states which we represent with integers in ${\mathbb Z}_d$ (the integers modulo $d$).
In discrete-time Markov chains we have a sequence of steps (times) $t=1,2,...$. At each step we have a probability vector $(x_0,...,x_{d-1})$ that gives probabilities for the system to be in the states $0,...,d-1$, correspondingly.

A `discrete path' $\wp(1,n)$ of the system between $t=1$ and $t=n$ is a sequence of $n$ integers in ${\mathbb Z}_d$:
\begin{eqnarray}\label{18}
{\wp}(1,n)=(a_1,...,a_n)\in ({\mathbb Z}_d)^n.
\end{eqnarray}
It shows that the system is in the state $\Sigma=a_1$ at $t=1$, it is in the state $\Sigma=a_2$ at $t=2$, etc. There are $d^n$ paths between $t=1$ and $t=n$.

We assume that at time $t=1$ the probability vector is 
\begin{eqnarray}\label{18}
{\bf x}^{(1)}=(x_0^{(1)},...,x_{d-1}^{(1)}),
\end{eqnarray}
and that at time $t=2$ it becomes
 \begin{eqnarray}\label{hh}
{\bf x}^{(2)}={\bf x}^{(1)}{\mathfrak D}_{12}.
\end{eqnarray}
In the literature on Markov chains the transition matrix ${\mathfrak D}_{12}$, is assumed to be a row Markov matrix.
We are interested in the special case where  ${\mathfrak D}_{12}$ is a doubly stochastic matrix.

${\mathfrak D}_{12}(a,b)$ is the probability that the state $\Sigma=a$ at $t=1$, will become the state $\Sigma=b$ at $t=2$.
Since the state $\Sigma=a$ at $t=1$ will become some state at $t=2$, it follows that $\sum _b{\mathfrak D}_{12}(a,b)=1$, i.e., ${\mathfrak D}_{12}$ is a row Markov matrix.
In a classical system there is no reason for the extra requirement $\sum _a{\mathfrak D}_{12}(a,b)=1$, which makes the matrix doubly stochastic.
We will see later that our system is quantum, and  in the time interval between $t=1$ and $t=2$ evolves with a unitary transformation.
In this case we have  quantum mechanical conservation of probabilities, which leads to the extra condition $\sum _a{\mathfrak D}_{12}(a,b)=1$.

Since at $t=1$ the system is in the state $\Sigma=a$ with probability ${x}^{(1)}_a$, it follows that
the probability that the system will follow the path $\wp(1,2)=(a,b)$, is
 \begin{eqnarray}
 {q}(a,b)={x}^{(1)}_a{\mathfrak D}_{12}(a,b);\;\;\;\sum_{b}q(a,b)=x_a^{(1)};\;\;\; \sum_a{q}(a,b)={x}^{(2)}_b;\;\;\;\sum_{a,b}q(a,b)=1.
 \end{eqnarray}

 We next assume that at time $t=3$ the probability vector becomes
 \begin{eqnarray}
{\bf x}^{(3)}={\bf x}^{(2)}{\mathfrak D}_{23}={\bf x}^{(1)}{\mathfrak D}_{13};\;\;\;{\mathfrak D}_{13}={\mathfrak D}_{12}{\mathfrak D}_{23}.
\end{eqnarray}
Here ${\mathfrak D}_{23}$ is a doubly stochastic transition matrix, and therefore ${\mathfrak D}_{13}$ is a doubly stochastic matrix (as a product of two doubly stochastic matrices).
For a time-dependent (non-homogeneous) Markov chain, ${\mathfrak D}_{23}$ is in general different from ${\mathfrak D}_{12}$. 
In this case the probability that the system will follow the path $\wp(1,3)=(a,b,c)$ is
\begin{eqnarray}
 q(a,b,c)={x}^{(1)}_a{\mathfrak D}_{12}(a,b){\mathfrak D}_{23}(b,c);\;\;\;\sum_c q(a,b,c)=q(a,b);\;\;\;\sum_{a,b} q(a,b,c)=x_c^{(3)}.
 \end{eqnarray}

We continue in a similar way and at time $t=s$ the probability vector becomes
 \begin{eqnarray}
{\bf x}^{(s)}={\bf x}^{(s-1)}{\mathfrak D}_{s-1,s}={\bf x}^{(1)}{\mathfrak D}_{1s};\;\;\;{\mathfrak D}_{1s}={\mathfrak D}_{12}...{\mathfrak D}_{s-1,s}.
\end{eqnarray}
Here ${\mathfrak D}_{1s}$ is a doubly stochastic transition matrix.
In this case the probability that the system will follow the path $\wp(1,s)=(a_1,...,a_{s})$ is
\begin{eqnarray}\label{path}
&&q(a_1,a_2,...,a_{s})={x}^{(1)}_{a_1}{\mathfrak D}_{12}(a_1,a_2){\mathfrak D}_{23}(a_2,a_3)...{\mathfrak D}_{s-1, s}(a_{s-1},a_{s});\nonumber\\
&&\sum _{a_1,...,a_s}q(a_1,a_2,...,a_{s})=1.
\end{eqnarray}
It is easily seen that
\begin{eqnarray}
\sum_{a_2,...,a_{s-1}}{\mathfrak D}_{12}(a_1,a_2){\mathfrak D}_{23}(a_2,a_3)...{\mathfrak D}_{s-1, s}(a_{s-1},a_{s})={\mathfrak D}_{1s}(a_1,a_{s}).
\end{eqnarray}
Using it we prove that the path probabilities obey the relations
\begin{eqnarray}
\sum _{a_2,...,a_s}q(a_1,a_2,...,a_{s})=x_{a_1}^{(1)};\;\;\;\sum _{a_1,...,a_{s-1}}q(a_1,a_2,...,a_{s})=x_{a_s}^{(s)}.
 \end{eqnarray}
They also obey the relations
 \begin{eqnarray}
\sum _{a_1,...,a_k}q(a_1,...,a_k,a_{k+1},...,a_{s})=q(a_{k+1},...,a_{s});\;\;\;k<s.
 \end{eqnarray}
 
The probability $q(A)$ that the path  $\wp(1,s)\in A$, where $A$ is a subset of $({\mathbb Z}_d)^s$ is 
\begin{eqnarray}
q(A)=\sum _{(a_1,...,a_s)\in A}q(a_1,a_2,...,a_{s});\;\;\;A\subseteq ({\mathbb Z}_d)^s
 \end{eqnarray}

We use the notation
\begin{eqnarray}\label{hhh}
{\mathfrak D}_{k\ell}={\mathfrak D}_{k,k+1}{\mathfrak D}_{k+1,k+2}...{\mathfrak D}_{\ell-1,\ell};\;\;\;k<\ell.
\end{eqnarray}
Then
\begin{eqnarray}
&&{\mathfrak D}_{k\ell}={\mathfrak D}_{kr}{\mathfrak D}_{r\ell};\;\;\;k<r<\ell
\end{eqnarray}
Also
\begin{eqnarray}
&&{\bf x}^{(s)}={\bf x}^{(k)}{\mathfrak D}_{ks};\;\;\;k<s
\end{eqnarray}
Knowledge of ${\bf x}^{(k)}$ alone determines all the future ${\bf x}^{(\ell)}$ with $k<\ell$ (knowledge of the past ${\bf x}^{(r)}$ with $r<k$ is not required for the calculation).
This `lack of memory' is the Markov property.
Given the `present', the `future' is independent of the `past'.
The present alone determines the future, and the history can be disregarded.

\subsection{Entropy of the probability vectors}\label{secQ1}

In this section we introduce entropic quantities associated with the probability vectors in the Markov chain. 
Using Eq.(\ref{11}) we get:
\begin{eqnarray}\label{E1}
{\bf u}\prec...\prec{\bf x}^{(s+1)}\prec {\bf x}^{(s)}\prec...\prec{\bf x}^{(1)},
\end{eqnarray}
where ${\bf u}$ is the uncertain vector in Eq.(\ref{5}),
or equivalently
\begin{eqnarray}\label{E2}
\log d\ge ...\ge E[{\bf x}^{(s+1)}]\ge E[{\bf x}^{(s)}]\ge...\ge E[{\bf x}^{(1)}]\ge 0.
\end{eqnarray}
The increase in the entropy is related to the fact that 
doubly stochastic matrices form a semigroup with respect to multiplication, and therefore the above process is in general irreversible.
In the very special case of a Markov chain with permutation matrices (which form the symmetric group), the process is reversible.

The 
\begin{eqnarray}
\Delta E(s,s+k)=E[{\bf x}^{(s+k)}]-E[{\bf x}^{(s)}]\ge 0
\end{eqnarray}
is the entropy production when the probability vector ${\bf x}^{(s)}$ becomes ${\bf x}^{(s+k)}$.
The following additivity relation holds:
\begin{eqnarray}
\Delta E(s,s+\ell)=\Delta E(s,s+k)+\Delta E(s+k,s+\ell);\;\;\;k\le \ell.
\end{eqnarray}

\subsection{Kullback-Leibler distance}\label{secQ2}
We quantify how close is the probability vector ${\bf x}^{(s)}$ to the uncertain vector ${\bf u}$, using the relative entropy 
\begin{eqnarray}
S({\bf x}^{(s)}||{\bf u})=\sum x_i^{(s)}\log\frac{x_i^{(s)}}{u_i}=\log d-E({\bf x}^{(s)});\;\;\;0\le S({\bf x}^{(s)}||{\bf u})\le \log d.
\end{eqnarray}
It is known as Kullback-Leibler distance between the two probability vectors\cite{RE1,RE2} (although mathematically it does not have the properties of a distance, for example $S({\bf x}^{(s)}||{\bf u})\ne S({\bf u}||{\bf x}^{(s)})$).
From Eq.(\ref{E2}) follows that as the step $s$ increases the  $S({\bf x}^{(s)}||{\bf u})$ decreases, which shows that ${\bf x}^{(s)}$ gets closer to the uncertain vector ${\bf u}$:
\begin{eqnarray}
0\le ...\le S({\bf x}^{(s+1)}||{\bf u})\le S({\bf x}^{(s)}||{\bf u})\le...\le S({\bf x}^{(1)}||{\bf u})\le \log d.
\end{eqnarray}
 If for some $s$ we get ${\bf x}^{(s)}={\bf u}$ then subsequently, for all $t>s$ we get ${\bf x}^{(t)}={\bf u}$.

\subsection{Entropy of the discrete paths}\label{secQ3}
The probability that the system will follow the path $\wp(1,s)=(a_1,...,a_{s})$ is $q(a_1,a_2,...,a_{s})$ in Eq.(\ref{path}).
We define the following entropy related to these probabilities:
\begin{eqnarray}\label{36}
E_{\wp}(1,s)=-\frac{1}{s}\sum _{a_1,...,a_s}q(a_1,a_2,...,a_{s}) \log q(a_1,a_2,...,a_{s}).
\end{eqnarray}
The number of paths is $d^s$ and we use the normalization factor $\frac{1}{s}$ so that the maximum value of $E_{\wp}(1,s)$ is $\log d$:
\begin{eqnarray}
0\le E_{\wp}(1,s)\le \log d.
\end{eqnarray}
$E_{\wp}(1,s)$ quantifies the randomness in the path of the system.
Similar quantities are used in the context of dynamical systems \cite{D1,D2}, and here we use them in the first instance for Markov chains and later for quantum systems.
It is easily seen that
\begin{eqnarray}
&&E_{\wp}(1,s)=\frac{1}{s}\left \{E[{\bf x}^{(1)}]+\sum_{a_1}x_{a_1}^{(1)}{\cal E}(a_1)\right \}\nonumber\\
&&{\cal E}(a_1)=-\sum _{a_2,...,a_s}[{\mathfrak D}_{12}(a_1,a_2)...{\mathfrak D}_{s-1, s}(a_{s-1},a_{s})]
\log [{\mathfrak D}_{12}(a_1,a_2)...{\mathfrak D}_{s-1, s}(a_{s-1},a_{s})]
\end{eqnarray}

\subsection{Ergodic Markov chain with doubly stochastic matrices}\label{sec89}

There are various definitions of ergodic Markov chains in the literature, for the general case that the transition matrices are row Markov matrices\cite{E1,E2,E3,E4,E5}.
Some of these definitions  are weaker than others.
The general idea is that in the large $n$ limit, all rows in the transition matrices ${\mathfrak D}_{k,k+n}(a,b)$ should be `almost equal' to a `stationary probability vector' (which does not depend on the row number $a$).

In our Markov chains the transition matrices are doubly stochastic, and it follows that the stationary probability vector can only be the most uncertain probability vector ${\bf u}$ in Eq.(\ref{5}).
$\bf u$ is a stationary probability vector for any doubly stochastic matrix (Eq.(\ref{6})).
In this case the general idea is that in the large $n$ limit, the transition matrices ${\mathfrak D}_{1,n}(a,b)$ should be `almost equal' to ${\mathfrak U}$ (in Eq.(\ref{UUU})).

A Markov chain with doubly stochastic matrices is ergodic if 
 \begin{eqnarray}\label{44}
\lim _{n\rightarrow\infty}{\mathfrak D}_{1,n}(a,b)=\frac{1}{d}
\end{eqnarray}
In this case for every infinitesimal $\epsilon >0$, there exist $n_0$ such that for all $n>n_0$
 \begin{eqnarray}
\frac{1}{d}-\epsilon\le {\mathfrak D}_{1,n}(a,b)\le\frac{1}{d}+\epsilon.
\end{eqnarray}
We note that:
\begin{itemize}
\item
for every pair of states $(a,b)$ there exists a non-zero probability that the state $\Sigma =a$ at the step $t=1$, will be transformed to  the state $\Sigma =b$ at some step $t=n$.
This is called irreducibility.
\item
The probability ${\mathfrak D}_{1,n}(a,a)$ is non-zero for all states $\Sigma=a$ and for all $n>n_0$. This is called aperiodicity.
\end{itemize}

From Eq.(\ref{44}) follows that the probability for the  path $\wp(1,n)=(a_1,...,a_{n_0},...,a_n)$ is 
\begin{eqnarray}\label{path}
&&A\left (\frac{1}{d}-\epsilon \right )^{n-n_0}\le q(a_1,,...,a_{n_0},...,a_n)\le A\left (\frac{1}{d}+\epsilon \right )^{n-n_0}\nonumber\\
&&A={x}^{(1)}_{a_1}{\mathfrak D}_{12}(a_1,a_2)...{\mathfrak D}_{n_0-1, n_0}(a_{n_0-1},a_{n_0})
\end{eqnarray}
and we find that
\begin{eqnarray}
\lim _{n\rightarrow \infty}\left [q(a_1,...,a_n)-\frac{1}{d^n}\right ]=0.
\end{eqnarray}
This means that in the ergodic Markov chain and in the large $n$ limit, the system follows all paths $\wp(1,n)$ with the same probability $\frac{1}{d^n}$.
Consequently the entropy of the paths takes the largest value $\log d$:
\begin{eqnarray}
\lim _{n\rightarrow \infty}E_{\wp}(1,n)=\log d.
\end{eqnarray}

\begin{example}
We consider a homogeneous Markov chain with
\begin{eqnarray}
{\mathfrak D}_{12}={\mathfrak D}_{23}=...={\mathfrak D}_{s-1,s}={\mathfrak U}.
\end{eqnarray}
According to our definition this is an ergodic Markov chain.
We consider two initial probability vectors.
\begin{itemize}
\item
The first one is ${\bf x}^{(1)}={\bf u}$. Then 
\begin{eqnarray}
q(a_1,a_2,...,a_{s})=\frac{1}{d^s};\;\;\;E_{\wp}(1,s)= \log d.
\end{eqnarray}
The system follows all paths with the same probability $\frac{1}{d^s}$.
\item
The second initial probability vector is ${\bf x}^{(1)}=(1,0,...,0)$. Then 
\begin{eqnarray}
q(a_1,a_2,...,a_{s})=\frac{\delta(a_1,0)}{d^{s-1}};\;\;\;
E_{\wp}(1,s)= \frac{s-1}{s}\log d.
\end{eqnarray}
In the large $s$ limit the system follows all paths with the same probability:
\begin{eqnarray}
q(a_1,a_2,...,a_{s})\rightarrow \frac{1}{d^{s}}\;\;\;\;E_{\wp}(1,s)\;\rightarrow\;\log d.
\end{eqnarray}
\end{itemize}
\end{example}
\begin{example}
We consider a homogeneous Markov chain case with
\begin{eqnarray}
{\mathfrak D}_{12}={\mathfrak D}_{23}=...={\mathfrak D}.
\end{eqnarray}
Therefore
\begin{eqnarray}
{\bf x}^{(s+1)}={\bf x}^{(1)}{\mathfrak D}^s.
\end{eqnarray}
We next assume that all eigenvalues of ${\mathfrak D}$ apart from the eigenvalue $e_0=1$,
have absolute value  $|e_i|<1$.
Then as $s\rightarrow\;\infty$ the ${\mathfrak D}^s$
has the eigenvalues $e_0^s=1$, and $e_i^s$ with $|e_i^s|\rightarrow\;0$.
Using Eq.(\ref{8}), we get
\begin{eqnarray}\label{A1}
\lim _{s\rightarrow\infty}{\mathfrak D}^s(a,b)=\frac{1}{d}.
\end{eqnarray}
Therefore this is ergodic Markov chain.
The $e_{\rm max}=\max \{|e_1|,...,|e_{d-1}|\}$ defines how quickly we will reach the limit.
The $(e_{\rm max})^s$ should be close to zero, and this leads to
\begin{eqnarray}
s\gg \frac{1}{-\log e_{\rm max}}.
\end{eqnarray}

\end{example}

\subsection{Universal convex polytopes}\label{sec450}

In this section we define  convex polytopes related to the probability vectors.
They are for Markov chains with doubly stochastic matrices, and for this reason they have not been studied in the literature
that uses general row Markov matrices.

Below we sometimes represent a probability vector with its first $d-1$ independent components (the last component is not independent).
This is indicated with a `hat' in the notation:
\begin{eqnarray}
{\bf x}=(x_0,...,x_{d-1})\;\rightarrow\;\widehat{\bf x}=(x_0,...,x_{d-2});\;\;\;\sum _{i=0}^{d-2}x_i\le 1.
\end{eqnarray}
For example, the probability vector ${\bf x}=(0,...,0,1)$ is represented with $\widehat{\bf x}=(0,...,0)$.

We define the simplex $\Delta _{d-1}$ 
\begin{eqnarray}
\Delta_{d-1}=\left \{(t_0,...,t_{d-2})\; |\;\sum _{i=0}^{d-2}t_i\le 1\right \},
\end{eqnarray}
which contains all probability vectors $\widehat{\bf x}$.
It is a $(d-1)$-dimensional convex polytope, with the $d$ vertices 
\begin{eqnarray}
{\rm vert}(\Delta_{d-1})=\left \{(0,...,0),(0,...,0,1), ...,(1,0,...,0)\right \}.
\end{eqnarray} 

We consider a Markov chain with the probability vector ${\bf x}^{(1)}$ (in Eq.(\ref{18})) at time $t=1$.
We define the 
\begin{eqnarray}\label{tt}
{\bf x}^{(1)}_\pi={\bf x}^{(1)}M_\pi=\left (x_{\pi^{-1}(0)}^{(1)},...,x_{\pi^{-1}(d-1)}^{(1)}\right );\;\;\;E({\bf x}^{(1)}_\pi)=E({\bf x}^{(1)}).
\end{eqnarray}
The components of ${\bf x}^{(1)}_\pi$ are a permutation of the components of ${\bf x}^{(1)}$.
There are $d!$ such vectors, but some of them might be equal to each other. 

We define the convex polytope ${\cal A}[{\bf x}^{(1)}]$ that contains all the probability vectors that are convex combinations of the probability vectors $\widehat {\bf x}^{(1)} _\pi$
(represented as the hat notation indicates with their first $d-1$ components):
\begin{eqnarray}\label{680}
{\cal A}[{\bf x}^{(1)}]=\left \{\sum _{\pi}\lambda (\pi)\widehat {\bf x}^{(1)}_\pi ;\;\;\lambda (\pi) \ge 0;\;\;\;\sum _{\pi} \lambda(\pi )=1\right \}.
\end{eqnarray}
We note that:
\begin{itemize}
\item
${\cal A}[{\bf x}^{(1)}]$ is a universal convex polytope in the sense that it depends only on the probability vector ${\bf x}^{(1)}$.
It does not depend on the doubly stochastic matrices $D_t$ with $t>1$ that will be used in this Markov chain in the future.
In other words Markov chains with different doubly stochastic matrices, which start with the same probability vector ${\bf x}^{(1)}$, have the same convex
polytope ${\cal A}[{\bf x}^{(1)}]$.

\item

${\cal A}[{\bf x}^{(1)}]$ is a subset of $\Delta_{d-1}$.
Also ${\cal A}[{\bf x}^{(1)}]$ contains the vector $\widehat {\bf u}=(\frac{1}{d},...,\frac{1}{d})$.
We easily prove this by taking in Eq.(\ref{680}) only the $d$ permutations which are `shifts' 
\begin{eqnarray}
\left (x_{0}^{(1)},...,x_{d-2}^{(1)}\right ),\left (x_{1}^{(1)},...,x_{d-1}^{(1)}\right ),\left (x_{2}^{(1)},...,x_{d-1}^{(1)},x_{0}^{(1)}\right ),...,
\left (x_{d-1}^{(1)},x_{0}^{(1)},...,x_{d-3}^{(1)}\right )
\end{eqnarray}
with coefficients $1/d$. The vector $\widehat {\bf u}$ is only one point in the polytope and can be viewed as the zero-dimensional simplex $\Delta_0$.
Therefore
\begin{eqnarray}
\Delta_0\subseteq {\cal A}[{\bf x}^{(1)}]\subseteq \Delta_{d-1}.
\end{eqnarray}

\item
The minimum entropy of the probability vectors in ${\cal A}[{\bf x}^{(1)}]$  is $E({\bf x}^{(1)})$.
Indeed, the concavity property of the entropy shows that every probability vector ${\bf y}\in {\cal A}[{\bf x}^{(1)}]$ has entropy
\begin{eqnarray}
\log d\ge E({\bf y})=E\left(\sum _{\pi}\lambda (\pi) {\bf x}^{(1)}_\pi \right  ) \ge \sum _{\pi}\lambda (\pi) E({\bf x}^{(1)}_\pi) =E({\bf x}^{(1)}).
\end{eqnarray}

\end{itemize}

The importance of ${\cal A}[{\bf x}^{(1)}]$ lies in the fact that all future probability vectors in the Markov chain belong to ${\cal A}[{\bf x}^{(1)}]$.

\begin{proposition}
In the above Markov chain:
\begin{itemize}
\item[(1)]
Every future probability vector ${\bf x}^{(s+1)}$ with $s>0$, belongs to the polytope ${\cal A}[{\bf x}^{(1)}]$ in Eq.(\ref{680}).
\item[(2)]
every probability vector ${\bf y}\in {\cal A}[{\bf x}^{(1)}]$ is a potential future probability vector, in the sense that there exists a doubly stochastic matrix ${\mathfrak D}$ such that
${\bf y}={\mathfrak D}{\bf x}^{(1)}$.
\end{itemize}
\end{proposition}
\begin{proof}
\begin{itemize}
\item[(1)]
Using the Birkhoff-von Neumann expansion
\begin{eqnarray}\label{123}
{\mathfrak D}_{1s}=\sum _{\pi}\alpha_{1s}(\pi)M_\pi
\end{eqnarray}
where $\alpha_{1s}(\pi)$ are probabilities, 
we express ${\bf x}^{(s+1)}$ as
 \begin{eqnarray}
{\bf x}^{(s+1)}={\bf x}^{(1)}{\mathfrak D}_{1s}=\sum _{\pi}\alpha_{1s}(\pi){\bf x}^{(1)}_\pi.
\end{eqnarray}
Therefore ${\bf x}^{(s+1)}$ belongs to the convex polytope ${\cal A}[{\bf x}^{(1)}]$.
\item[(2)]
Let ${\bf y}\in {\cal A}[{\bf x}^{(1)}]$.
Then by definition there exist probabilities $\lambda(\pi)$ such that ${\bf y}=\sum _{\pi}\lambda (\pi) {\bf x}^{(1)}_\pi$. But then
\begin{eqnarray}
{\bf y}={\bf x}^{(1)}{\mathfrak D};\;\;\;{\mathfrak D}=\sum _{\pi}\lambda (\pi)M_\pi.
\end{eqnarray}
\end{itemize}

\end{proof}

In analogy to ${\cal A}[{\bf x}^{(1)}]$ we define a convex polytope for every $s$:
\begin{eqnarray}\label{25}
\Delta _0\subseteq {\cal A}[{\bf x}^{(s)}]=\left \{\sum _{\pi}\lambda (\pi) \widehat {\bf x}^{(s)}_\pi ;\;\;\lambda (\pi) \ge 0;\;\;\;\sum _{\pi} \lambda(\pi )=1\right \}\subseteq \Delta_{d-1}.
\end{eqnarray}
${\cal A}[{\bf x}^{(s)}]$ contains all future probability vectors ${\bf x}^{(t)}$ with $t>s$, and therefore
\begin{eqnarray}\label{40}
\Delta_0\subseteq ...\subseteq {\cal A}[{\bf x}^{(s+1)}]\subseteq{\cal A}[{\bf x}^{(s)}]\subseteq ... \subseteq {\cal A}[{\bf x}^{(1)}]\subseteq \Delta_{d-1}.
\end{eqnarray}
As the step number $s$ increases the convex polytope shrinks, and its dimension might stay the same or it might decrease.
Also the minimum entropy of the probability vectors in the polytope increases, as $s$ increases.(Eq.(\ref{E1})).

${\cal A}[{\bf x}^{(s)}]$ is a universal convex polytope in the sense that it depends only on the probability vector ${\bf x}^{(s)}$.
It does not depend on the doubly stochastic matrices $D_t$ with $t>s$ that will be used in this Markov chain in the future.
 Of course the shrinking of this polytope in the future, depends on the doubly stochastic matrices $D_t$ with $t>s$.

\begin{figure}
\centering
\includegraphics{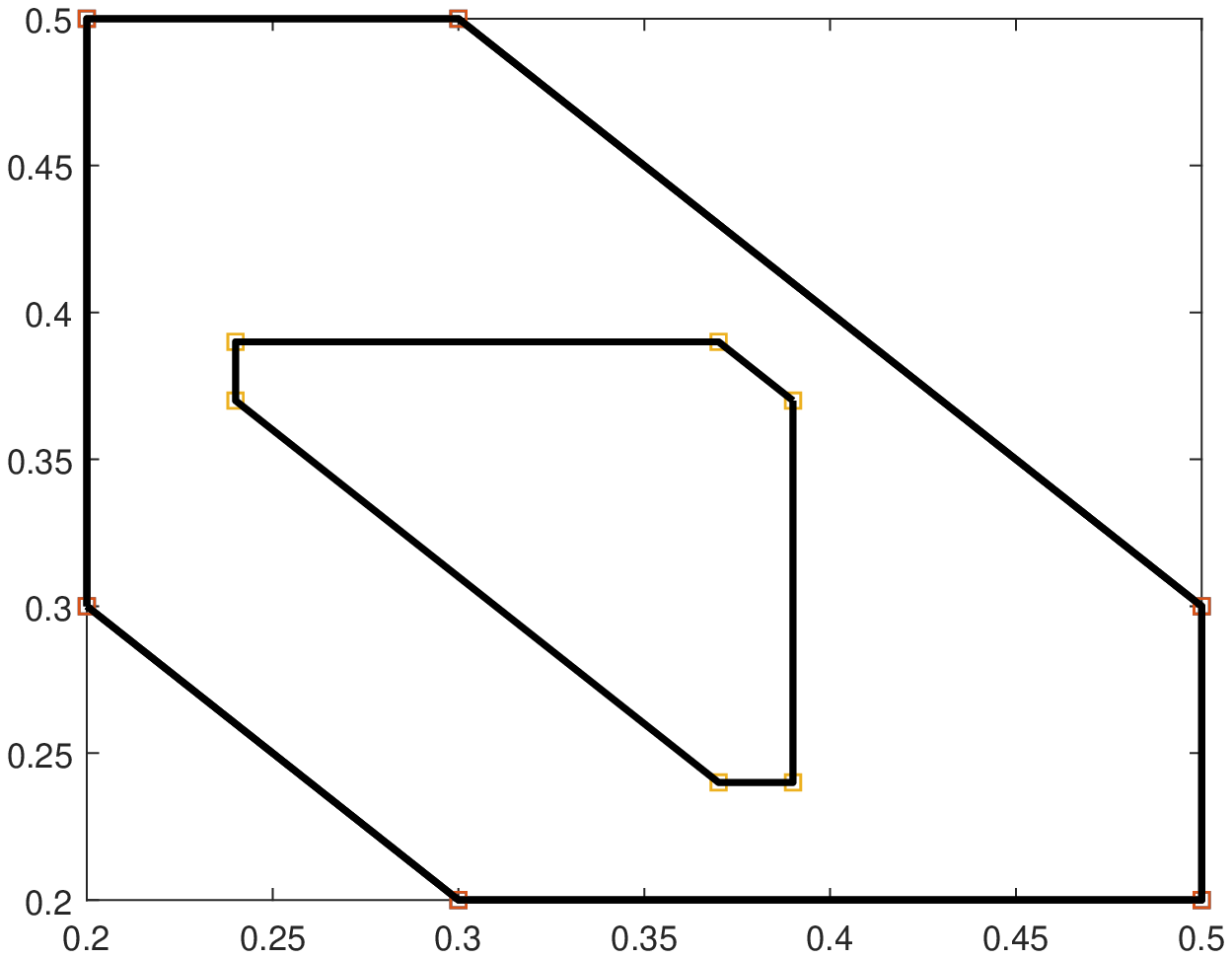}
\caption{The outer polytope is the convex polytope for a Markov chain, with the probability vector in Eq.(\ref{26}). 
With the doubly stochastic matrix in Eq.(\ref{45}), we get in
the next step the probability vector in Eq.(\ref{27}), and then the convex polytope shrinks into the inner one.}
\label{FF1}
\end{figure}

\begin{example}
For $d=3$, we consider a Markov chain with the probability vector
\begin{eqnarray}\label{26}
{\bf x}^{(1)}=(0.2, 0.3, 0.5),
\end{eqnarray}
at $t=1$. We have six permutations of the vector ${\bf x}^{(1)}$.
We write the six vectors  $\widehat {\bf x}^{(1)}_\pi$ as the matrix
\begin{eqnarray}
M=\begin{pmatrix}
0.2&0.3\\
0.2&0.5\\
0.3&0.2\\
0.5&0.2\\
0.3&0.5\\
0.5&0.3
\end{pmatrix}.
\end{eqnarray}
We use this matrix with the program `convhull' in MATLAB and we get the outer polytope in Fig.\ref{FF1}, which is $2$-dimensional and has six vertices.
The polytope is universal in the sense that all future probability vectors (their first two components) belong to this polytope, regardless of the doubly stochastic matrices that will be used  in this Markov chain. 
For example, we see from Fig.\ref{FF1} that in the future we will never get the probability vector $(0.4,0.5,0.1)$.

For the next step, we consider the doubly stochastic matrix
\begin{eqnarray}\label{45}
{\mathfrak D}_{12}=\begin{pmatrix}
0.1&0.3&0.6\\
0.4&0.2&0.4\\
0.5&0.5&0
\end{pmatrix}
\end{eqnarray}
and we get the probability vector
\begin{eqnarray}\label{27}
{\bf x}^{(2)}={\bf x}^{(1)}{\mathfrak D}_{12}=(0.39, 0.37,0.24),
\end{eqnarray}
at $t=2$. We then repeat the above process and we find that the polytope shrinks into the inner polytope in Fig.\ref{FF1}.

The Kullback-Leibler distance of the probability vectors ${\bf x}^{(1)}$, ${\bf x}^{(2)}$ from ${\bf u}$ is
\begin{eqnarray}
S({\bf x}^{(1)}||{\bf u})=0.069;\;\;\;S({\bf x}^{(2)}||{\bf u})=0.021.
\end{eqnarray}
Here we used natural logarithms and therefore the result is in nats.

\end{example}

\begin{figure}
\centering
\includegraphics{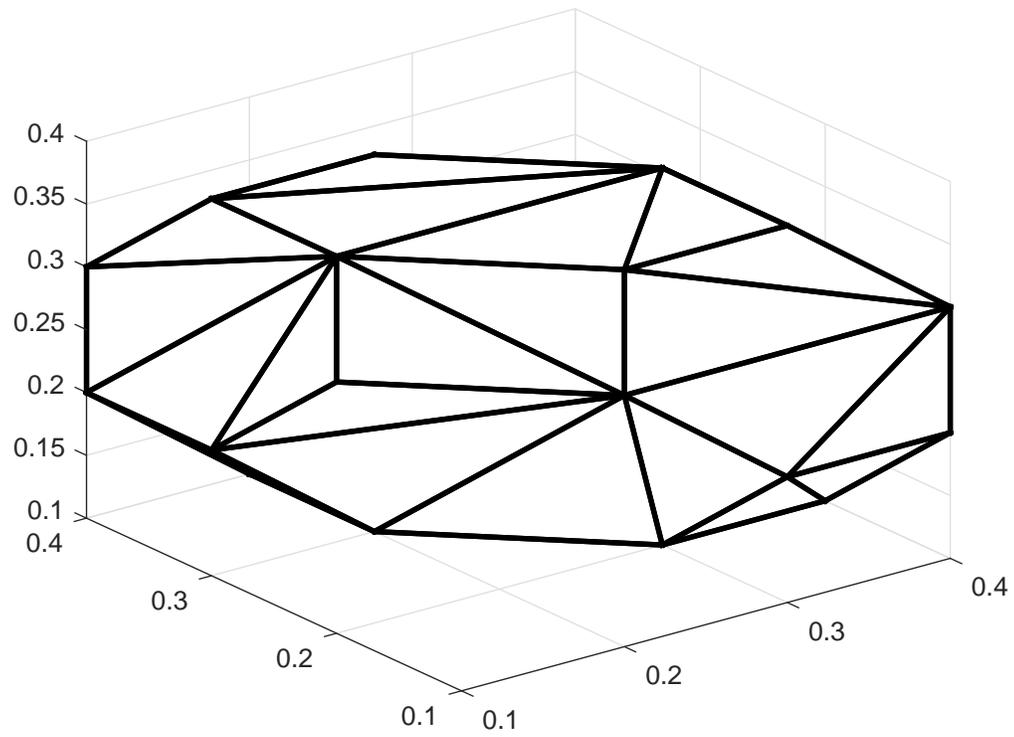}
\caption{The $3$-dimensional convex polytope for a Markov chain, with the probability vector in Eq.(\ref{26A}).}

\label{FF2}
\end{figure}

\begin{example}
For $d=4$, we consider 
a Markov chain, with probability vector
\begin{eqnarray}\label{26A}
{\bf x}^{(1)}=(0.1, 0.2, 0.3, 0.4),
\end{eqnarray}
at $t=1$. We have $24$ `permutation vectors' ${\bf x}^{(1)}_\pi$.
We use the same process as in the previous example and we get the $3$-dimensional convex polytope in Fig.\ref{FF2}.
\end{example}

\begin{example}
In a Markov chain at $t=1$ the probability vector is
${\bf x}^{(1)}=(1,0,...,0)$. There are only $d$ permutations of this vector which are different from each other, and which
are convexly independent vectors (none of them is a convex combination of the others).
Then  ${\cal A}[{\bf x}^{(1)}]$ is the $(d-1)$-simplex:
\begin{eqnarray}
{\cal A}[{\bf x}^{(1)}]=\Delta_{d-1}
\end{eqnarray}
We note that this contains all probability vectors, so in this particular example the convex polytope does not give  any useful information.
\end{example}
\begin{example}
If ${\bf x}^{(1)}={\bf u}$ (defined in Eq.(\ref{5})) all permutations ${\bf x}^{(1)}_\pi$ of this  vector are equal to each other. The convex polytope in this case is just one point:
\begin{eqnarray}\label{121}
{\cal A}[{\bf x}^{(1)}]=\Delta_0.
\end{eqnarray}

\end{example}

\section{A sequence of non-selective quantum measurements}\label{sec45}

\subsection{Quantum Mechanics is not a Markovian theory}\label{sec78}

We consider a quantum system (qudit) with variables in  ${\mathbb Z}_d$ (the integers modulo $d$). 
$H_d$ is the $d$-dimensional Hilbert space describing this system. 
$|f\rangle$ where $f\in {\mathbb Z}_d$, is an orthonormal basis in $H_d$ which we call basis of position states.
For later use, we also consider the Fourier transform
\begin{eqnarray}
{\cal F}=\frac{1}{\sqrt d}\sum _{f,g}\exp\left (\frac{2\pi fg}{d}\right)\ket{f}\bra{g}
\end{eqnarray}
Acting with ${\cal F}$ on the basis of position states, we get a dual basis that we call momentum states:
\begin{eqnarray}
\ket{f}_F={\cal F}\ket{f}
\end{eqnarray}

We assume that initially the system is described by a density matrix $\rho$ with elements $\rho(f,g)=\bra{f}\rho\ket{g}$ (in the basis of position states).
The density matrix evolves in time with a unitary matrix $V$ into $\sigma=V^\dagger \rho V$ which has matrix elements
\begin{eqnarray}
\sigma(f,g)=\sum _h[V(h,f)]^*\rho (h,k) V(k,g)
\end{eqnarray}
If the density matrix $\rho$ is diagonal ($\rho(f,g)=0$ for $f\ne g$) then the diagonal elements of $\sigma$ are given by
\begin{eqnarray}\label{gg}
\sigma(f,f)=\sum _h \rho (h,h) {\mathfrak D}(h,f);\;\;{\mathfrak D}(h,f)=|V(h,f)|^2.
\end{eqnarray}
Since $V$ is unitary matrix, it follows that ${\mathfrak D}(h,f)$ is a doubly stochastic matrix and then Eq.(\ref{gg}) is analogous to Eq.(\ref{hh}) in Markov chains.
But $\sigma$ also has the non-diagonal elements 
\begin{eqnarray}
\sigma(f,g)=\sum _h [V(h,f)]^*\rho (h,h) V(h,g);\;\;\;f\ne g.
\end{eqnarray}
It is seen that if the initial density matrix $\rho$ is diagonal, time evolution gives a density matrix $\sigma$ which has diagonal elements given by the Markov relation
in Eq.(\ref{gg}), but which also has non-diagonal elements $\sigma(f,g)$.
Non-diagonal elements in the density matrix are intimately related to the superposition principle, and are at the heart of Quantum Mechanics.
 It is therefore clear that Quantum Mechanics is not a Markovian theory (see also \cite{GI}).
 
 Quantum evolution is a reversible process associated with a group of unitary transformations, while a Markov chain is an irreversible process associated with a semigroup.
 A generalisation of Markov theory called quantum Markov theory,  suitable for the description of quantum systems and their interaction with the environment, has been studied in \cite{AC1,AC2,AC3,AC4,AC5}.
 
Quantum systems can be described with Markov chain if we have a mechanism of destroying the non-diagonal elements and introduce irreversibility.
We will do this stroboscopically with a sequence of non-selective measurements. 
Our formalism below is a Markovian stroboscopic description of the system, immediately after the non-selective measurements.
In the time intervals between the measurements the system acquires off-diagonal elements in the density matrix, and cannot be described in a Markovian way.

\subsection{Non-selective measurements on qudits}

We consider the quantum system (qudit) described in section \ref{sec78}.
Let ${\mathfrak P}_f$ be the orthogonal projectors of rank one:
\begin{eqnarray}\label{33}
{\mathfrak P}_f=\ket{f}\bra{f};\;\;\;\sum _{f=0}^{d-1}{\mathfrak P}_f={\bf 1}.
\end{eqnarray}
Measurements with ${\mathfrak P}_f$ at $t=1$ on a system described with the density matrix $\rho$,  will give `yes' with probability 
\begin{eqnarray}
x_f^{(1)}={\rm Tr}[\rho {\mathfrak P}_f]=\rho(f,f);\;\;\;\sum _{f=0}^{d-1}x_f^{(1)}=1.
\end{eqnarray}
The probability vector
\begin{eqnarray}\label{a1}
{\bf x}^{(1)}=(x_0^{(1)},...,x_{d-1}^{(1)})
\end{eqnarray}
describes the outcome from the measurements ${\mathfrak P}_f$ (with $f=0,...,d-1$).

In the case of a non-selective measurement (where we do not look at the outcome of the measurement) the system will be described after the measurement, with the density matrix
\begin{eqnarray}
\rho _1=\sum _f{\mathfrak P}_f\rho{\mathfrak P}_f.
\end{eqnarray}
The non-selective measurement destroys the off-diagonal terms ${\mathfrak P}_f\rho{\mathfrak P}_g$ with $f\ne g$.
For orthogonal projectors of rank one (like the ${\mathfrak P}_f$), this can be also be written as
\begin{eqnarray}\label{az}
\rho _1=\sum _f x_f^{(1)}{\mathfrak P}_f=\sum _f\rho(f,f)\ket{f}\bra{f}.
\end{eqnarray}
In the basis of position states, $\rho_1$ is a diagonal matrix which has the probability vector in Eq.(\ref{a1}) in the diagonal:
\begin{eqnarray}\label{az}
\rho _1={\rm diag}(x_0^{(1)},...,x_{d-1}^{(1)}).
\end{eqnarray}

The von Neumann entropy of a density matrix $\rho$ is $E(\rho)=-{\rm Tr}(\rho \log \rho)$.
Since $\rho_1$ is a diagonal matrix, we get
\begin{eqnarray}
E(\rho _1)=E({\bf x}^{(1)}).
\end{eqnarray}

Diagonal density matrices can be represented by the probability vectors in their diagonals, and in analogy to ${\cal A}[{\bf x}^{(1)}]$ 
we define the convex polytope ${\cal B}(\rho_1)$ of these diagonal density matrices. 
${\cal A}[{\bf x}^{(1)}]$ has been defined in terms of probability vectors, and 
${\cal B}(\rho_1)$ in terms of diagonal density matrices that have these probability vectors in their diagonals. 
Then
\begin{eqnarray}
\Delta _0\subseteq {\cal B}(\rho_1)\subseteq \Delta_{d-1}.
\end{eqnarray}
Here $\Delta_{d-1}$ contains all diagonal $d\times d$ density matrices, and $\Delta _0$ is a zero-dimensional simplex (one point) related to the density matrix $\rho=\frac{1}{d}{\bf 1}$.

\subsection{A sequence of non-selective measurements as a Markov chain}

We next assume that this system evolves between $t=1$ and $t=2$ with the unitary matrix $V_{12}$ (which has matrix elements $V_{12} (f,g)$ in the basis of position states).
At $t=2$ we get the density matrix
\begin{eqnarray}\label{nb}
\widetilde \rho _2=\sum _{f}x_f^{(1)}{\mathfrak P}_f^{(2)};\;\;\;{\mathfrak P}_f^{(2)}=V_{12}^\dagger {\mathfrak P}_fV_{12}.
\end{eqnarray}
On the density matrix ${\widetilde \rho}_2$, we perform again a non-selective measurement with different  projectors which we write as
$W_{2}{\mathfrak P}_fW_{2}^\dagger$, where $W_{2}$ is a unitary matrix.
We then get
\begin{eqnarray}\label{nb10}
&&\rho _2=\sum _{f}{\rm Tr}(\widetilde \rho _2W_2{\mathfrak P}_fW_2^\dagger)W_{2}{\mathfrak P}_fW_{2}^\dagger=\sum _{f,g}x_g^{(1)}{\mathfrak D}_{12}(g,f)W_2{\mathfrak P}_fW_2^\dagger;\;\;\;g,f=0,...,d-1\nonumber\\
&&{\mathfrak D}_{12}(g,f)={\rm Tr}({\mathfrak P}_g^{(2)}W_2{\mathfrak P}_fW_2^\dagger)=|(V_{12}W_2)(g,f)|^2.
\end{eqnarray}
Since $V_{12}W_2$ is a unitary matrix, ${\mathfrak D}_{12}(g,f)$ is a  doubly stochastic transition matrix. 
The fact that our system evolves quantum mechanically between successive measurements with a unitary transformation (which in turn is related to conservation of probabilities), leads to doubly stochastic transition matrices.
A system which is classical between $t=1$ and $t=2$, will be described with general row Markov transition matrices.

We rewrite Eq.(\ref{nb10}) as
\begin{eqnarray}\label{nb2}
\rho _2=\sum _{f}{x}_f^{(2)}W_2{\mathfrak P}_fW_2^\dagger;\;\;\;{x}_f^{(2)}=\sum _{g}x_g^{(1)}{\mathfrak D}_{12}(g,f)=\sum _{g}\rho(g,g){\mathfrak D}_{12}(g,f).
\end{eqnarray}
The ${\bf x}^{(2)}={\bf x}^{(1)}{\mathfrak D}_{12}$ is also a probability vector related to the second measurement.
$\rho_2$ in the basis $W_2\ket{f}$ is the diagonal matrix:
\begin{eqnarray}
\rho _2={\rm diag}(x_0^{(2)},...,x_{d-1}^{(2)}).
\end{eqnarray}

We next assume that this system evolves between $t=2$ and $t=3$ with unitary transformations $V_{23}$ and we get the density matrix
\begin{eqnarray}\label{nb}
\widetilde \rho _3=\sum _{f}{x}_f^{(2)}{\mathfrak P}_f^{(3)};\;\;\;{\mathfrak P}_f^{(3)}=V_{23}^\dagger W_2{\mathfrak P}_fW_2^\dagger V_{23}.
\end{eqnarray}
On the density matrix ${\widetilde \rho}_3$, we perform again a non-selective measurement with different projectors $W_3{\mathfrak P}_fW_3^\dagger$ where $W_3$ is a unitary matrix, and we get
\begin{eqnarray}\label{nb3}
&&\rho _3=\sum _{f}{\rm Tr}(\widetilde \rho _3W_3{\mathfrak P}_fW_3^\dagger)W_3{\mathfrak P}_fW_3^\dagger=\sum _{f,g}{x}_g^{(2)}{\mathfrak D}_{23}(g,f)W_3{\mathfrak P}_fW_3^\dagger=\sum _{f,}{x}_f^{(3)}W_3{\mathfrak P}_fW_3^\dagger
\nonumber\\
&&{\mathfrak D}_{23}(g,f)={\rm Tr}( {\mathfrak P}_{g}^{(3)}W_3{\mathfrak P}_fW_3^\dagger)=|(W_2^\dagger V_{23}W_3)(g,f)|^2\nonumber\\
&&{\bf x}^{(3)}={\bf x}^{(2)}{\mathfrak D}_{23}
\end{eqnarray}
${\mathfrak D}_{23}(g,f)$ is a doubly stochastic matrix. We rewrite Eq.(\ref{nb3}) as
\begin{eqnarray}\label{nb4}
&&\rho _3=\sum _{f,g,h}{x}_h^{(1)}{\mathfrak D}_{12}(h,g){\mathfrak D}_{23}(g,f)W_3{\mathfrak P}_fW_3^\dagger
=\sum _{f,h}{x}_h^{(1)}{\mathfrak D}_{13}(h,f)W_3{\mathfrak P}_fW_3^\dagger\nonumber\\
&&{\mathfrak D}_{13}(h,f)=\sum _{g}{\mathfrak D}_{12}(h,g){\mathfrak D}_{23}(g,f)=\sum _{g}|(V_{12}W_2)(h,g)|^2|(W_2^\dagger V_{23}W_3)(g,f)|^2
\end{eqnarray}
Here ${\mathfrak D}_{13}={\mathfrak D}_{12}{\mathfrak D}_{23}$  is the product of two doubly stochastic matrices, and therefore it is a doubly stochastic matrix.
$\rho_3$ is the diagonal matrix:
\begin{eqnarray}
\rho _3={\rm diag}(x_0^{(3)},...,x_{d-1}^{(3)}).
\end{eqnarray}

We repeat this process many times. It is a sequence of non-selective measurements with unitary transformations that describe time evolution, between them.
The unitary transformations  are in general different in each step. 
The projectors related to the measurements are also in general different in each step.
On the density matrix ${\widetilde \rho}_s$, we perform a non-selective measurement with the projectors $W_s{\mathfrak P}_fW_s^\dagger$ where $W_s$ is a unitary matrix
Schematically we describe this process as follows:
\begin{eqnarray}
&&\rho  \;\;\overset {\rm meas}\longrightarrow\;\; \rho _1\;\;\overset {V_{12}}\longrightarrow\;\;{\widetilde \rho }_2=V_{12}^\dagger \rho _1 V_{12}
\;\;\overset {\rm meas}\longrightarrow\;\; \rho _2
\;\;\overset {V_{23}}\longrightarrow\;\;{\widetilde \rho }_3=V_{23}^\dagger \rho _2 V_{23}\;\;\overset {\rm meas}\longrightarrow\;\;\rho _3
\overset {V_{34}}\longrightarrow...
\;\;\nonumber\\&& \overset {V_{s,s+1}}\longrightarrow \;\;{\widetilde \rho }_{s+1}=V_{s,s+1}^\dagger \rho _{s,s+1} V_{s+1}
\;\;\overset {\rm meas}\longrightarrow\;\;\rho_{s+1}\longrightarrow ...
\end{eqnarray}
Here 
\begin{eqnarray}\label{nb5}
&&\rho _{s+1}=\sum _{f,g}{x}_f^{(s+1)}W_{s+1}{\mathfrak P}_fW_{s+1}^\dagger=\sum _{f,g}{x}_g^{(s)}{\mathfrak D}_{s,s+1}(g,f)W_{s+1}{\mathfrak P}_fW_{s+1}^\dagger\nonumber\\
&&{\mathfrak D}_{s,s+1}(g,f)=|(W_s^\dagger V_{s,s+1}W_{s+1})(g,f)|^2
\end{eqnarray}
It is seen that
\begin{eqnarray}
&&{\bf x}^{(s+1)}={\bf x}^{(s)}{\mathfrak D}_{s,s+1}={\bf x}^{(k)}{\mathfrak D}_{k,s+1};\;\;\;k<s+1\nonumber\\
&&\rho _{s+1}={\rm diag}(x_0^{(s+1)},...,x_{d-1}^{(s+1)}).
\end{eqnarray}
where we use the notation of Eq.(\ref{hhh}).

Entropic inequalities in Eqs(\ref{E1}),(\ref{E2}) hold here also. We get the `entropic staircase':
\begin{eqnarray}
\log d\ge ...\ge E(\rho_{s+1})\ge E(\widetilde \rho_{s+1})=E(\rho_s)\ge...\ge E(\widetilde \rho_2)=E(\rho_1)\ge E(\rho).
\end{eqnarray}
Unitary time evolution is a reversible process that keeps the entropy constant.
The increases in the entropy are related to the measurements, and make the process irreversible.

We also define probabilities for discrete paths (as in Eq.(\ref{path})) by looking stroboscopically at the system at the times after the measurements.
The corresponding entropy is given by Eq.(\ref{36}).

We note that using different projectors in each step, simply changes the evolution matrix (e.g., from $V_{s,s+1}$ to $W_s^\dagger V_{s,s+1}W_{s+1}$ in Eq.(\ref{nb5})).
In the rest of the paper we assume that all projectors used in the measurements are the same, i.e., that $W_s={\bf 1}$.

\subsection{Universal convex polytopes}

In analogy to ${\cal B}(\rho_{1})$ we define a convex polytope ${\cal B}(\rho_s)$ for every $s$.
They are analogous to the polytopes ${\cal A}[{\bf x}^{(s)}]$, and we get here results analogous to those in section \ref{sec450}:
\begin{itemize}
\item
${\cal B}(\rho_s)$ is universal in the sense that it depends only on $\rho_s$, and it does not depend on the future time evolution matrices $V_{t}$ with $t>s$.
\item
${\cal B}(\rho_s)$ contains all future diagonal density matrices $\rho_{t}$ with $t>s$ (after each non-selective measurement).
\item
The minimum entropy of all density matrices in ${\cal B}(\rho_s)$ is $E(\rho_s)$.
\item
As the discrete time $s$ increases, the convex polytope shrinks:
\begin{eqnarray}\label{d2}
\Delta_0\subseteq...\subseteq{\cal B}(\rho_{s+1})\subseteq {\cal B}(\rho_s)\subseteq...\subseteq {\cal B}(\rho_1)\subseteq\Delta_{d-1},
\end{eqnarray}
and the minimum entropy of the density matrices in it increases.
\end{itemize}

\subsection{Destruction of the non-diagonal terms with non-selective measurements and the Markov property}\label{sec79}

We have explained in section \ref{sec78}
that time evolution in quantum mechanics is non-Markovian because of the non-diagonal terms.
The process between two non-selective measurements  can be described as a Markov chain, because 
the non-selective measurements destroy the non-diagonal elements.

In order to see this in a different way we compare and contrast the process
\begin{eqnarray}
&&\rho \overset {\rm meas} \longrightarrow\;\;\rho _1\;\;\overset {V_{12}}\longrightarrow\;\;{\widetilde \rho }_2=V_{12}^\dagger \rho _1 V_{12}
\;\;\overset {\rm meas}\longrightarrow\;\; \rho _2
\;\;\overset {V_{23}}\longrightarrow\;\;{\widetilde \rho }_3=V_{23}^\dagger \rho _2 V_{23}\;\;\overset {\rm meas}\longrightarrow\;\;\rho _3
\end{eqnarray}
with the case where the second non-selective measurement does not take place:
\begin{eqnarray}
&&\rho \overset {\rm meas} \longrightarrow\;\;\rho _1\;\;\overset {V_{12}}\longrightarrow\;\;{\widetilde \rho }_2=V_{12}^\dagger \rho _1 V_{12}
\;\;\longrightarrow\;\; \sigma _2={\widetilde \rho }_2
\;\;\overset {V_{23}}\longrightarrow\;\;{\widetilde \sigma }_3=V_{23}^\dagger \sigma _2 V_{23}\;\;\overset {\rm meas}\longrightarrow\;\;\sigma _3
\end{eqnarray}
In this case
\begin{eqnarray}\label{nb40}
&&\sigma _3=\sum _{f,h}{x}_h^{(1)}{\breve {\mathfrak D}_{12}(h,f)}{\mathfrak P}_f\nonumber\\
&&{\breve {\mathfrak D}_{12}(h,f)}=|(V_{12}V_{23})(h,f)|^2=\left |\sum _{g}V_{12}(h,g)V_{23}(g,f)\right |^2
\end{eqnarray}
${\breve {\mathfrak D}_{12}(h,f)}$ is a doubly stochastic matrix.
We compare Eqs(\ref{nb4}),(\ref{nb40}) and we note that 
\begin{eqnarray}
 {\breve {\mathfrak D}_{12}(h,f)}= {\mathfrak D}_{12}(h,f)+\sum _{g\ne g^\prime}V_{12}(h,g)V_{23}(g,f)V_{12}(h,g^\prime)V_{23}(g^\prime,f)
\end{eqnarray}
The difference between ${\breve {\mathfrak D}_{12}(h,f)}$ and ${\mathfrak D}_{12}(h,f)$ is non-diagonal terms which are absent in the latter case because the non-selective measurement destroys them.

\section{Special cases}\label{secww}

In this section we consider special cases of the general theory in section \ref{sec45}.

\subsection{Initial state is a position state}
We assume that the initial state of the system is the position state $\ket{r}$.
We perform a non-selective measurement with the projectors ${\mathfrak P}_f$ and we get the probability vector ${\bf x}^{(1)}$: 
\begin{eqnarray}
{x}_f^{(1)}=\delta(r,f).
\end{eqnarray}
After the non-selective measurement the system is described with the density matrix $\rho_1$: 
\begin{eqnarray}
\rho_1={\rm diag}(0,...,0,1,0,...,0).
\end{eqnarray}
Here the $(r,r)$-element of $\rho_1$ is one.
The corresponding entropy is 
\begin{eqnarray}
E(\rho _1)=E({\bf x}^{(1)})=0.
\end{eqnarray}
The corresponding convex polytope ${\cal B}(\rho_1)$ is the $(d-1)$-simplex:
\begin{eqnarray}
{\cal B}(\rho_1)=\Delta_{d-1}.
\end{eqnarray}

The system evolves in time with unitary transformations $V_{12}$ and we get the density matrix
\begin{eqnarray}\label{nb}
\widetilde \rho _2=V_{12}^\dagger\ket{r}\bra{r}V_{12}.
\end{eqnarray}
On the density matrix ${\widetilde \rho}_2$, we perform again a non-selective measurement with the same projectors ${\mathfrak P}_f$ and we get
\begin{eqnarray}\label{nb1}
{x}_f^{(2)}={\mathfrak D}_{12}(r,f)
;\;\;\;{\mathfrak D}_{12}(r,f)=|V_{12}(r,f)|^2
\end{eqnarray}
After the non-selective measurement the system is described with the diagonal density matrix $\rho_2$: 
\begin{eqnarray}
\rho _2={\rm diag} (|V_{12}(r,0)|^2,...,|V_{12}(r,d-1)|^2).
\end{eqnarray}

We continue in the same way and we get
\begin{eqnarray}
&&{\bf x}^{(s+1)}(g_{s+1})={\mathfrak D}_{1,s+1}(r,g_{s+1})\nonumber\\
&&{\mathfrak D}_{1,s+1}(r,g_{s+1})=\sum _{g_2,...,g_{s}}|V_{12}(r,g_2)|^2|V_{23}(g_2,g_3)|^2...|V_{s,s+1}(g_{s},g_{s+1})|^2.
\end{eqnarray}
After the non-selective measurement the system is described with the density matrix
\begin{eqnarray}
\rho _{s+1}={\rm diag}({\mathfrak D}_{1,s+1}(r,0),...,{\mathfrak D}_{1,s+1}(r,d-1)).
\end{eqnarray}

The probability that the system will follow the discrete path $\wp(1,s+1)=(g_1,g_2,...,g_{s+1})$ is
\begin{eqnarray}
q(g_1,g_2,...,g_{s})=\delta(r,g_1)|V_{12}(g_1,g_2)|^2|V_{23}(g_2,g_3)|^2...|V_{s,s+1}(g_{s},g_{s+1})|^2.
 \end{eqnarray}

\subsection{Initial state is a momentum state}

We assume that the initial state of the system is the momentum state $\ket{r}_F$.
We perform a non-selective measurement with the projectors ${\mathfrak P}_f$ and we get the probability vector: 
\begin{eqnarray}
{\bf x}^{(1)}={\bf u}=\left(\frac{1}{d},...,\frac{1}{d}\right)
\end{eqnarray}
After the non-selective measurement the system is described with the density matrix $\rho_1$: 
\begin{eqnarray}
\rho _{1}={\rm diag}\left(\frac{1}{d},...,\frac{1}{d}\right).
\end{eqnarray}

The corresponding convex polytope ${\cal B}(\rho_1)$ is the $0$-simplex (one point):
\begin{eqnarray}\label{100}
{\cal B}(\rho_1)=\Delta _0.
\end{eqnarray}
Then for any unitary evolution matrices $V_{12}, V_{23},...$, we get 
\begin{eqnarray}\label{P1}
&&{\bf x}^{(s+1)}={\bf u};\;\;\;\rho _{s+1}={\rm diag}\left(\frac{1}{d},...,\frac{1}{d}\right);\;\;\;{\cal B}(\rho_s)=\Delta _0,
\end{eqnarray}
for all $s=0,1,2,...$.Also
\begin{eqnarray}\label{P2}
E(\rho_{s+1})=E({\bf x}^{(s+1)})=\log d.
\end{eqnarray}

The probability that the system will follow the discrete path $\wp(1,s+1)=(g_1,g_2,...,g_{s+1})$ depends on the unitary evolution matrices $V_{12}, V_{23},...$, and it is given by
\begin{eqnarray}
q(g_1,g_2,...,g_{s})=\frac{1}{d}|V_{12}(g_1,g_2)|^2|V_{23}(g_2,g_3)|^2...|V_{s,s+1}(g_{s},g_{s+1})|^2.
 \end{eqnarray}
In this example all the probability vectors ${\bf x}^{(s+1)}$ are the same, but the probabilities of the various paths are different from each other.

\subsection{Initial state is $\rho=\frac{1}{d}{\bf 1}$}

We assume that the initial density matrix is 
\begin{eqnarray}
\rho=\frac{1}{d}{\bf 1}.
\end{eqnarray}
In this case we get results analogous to those in the previous example.

\subsection{Ergodic case}
We consider a quantum system with $d=3$ and Hamiltonian
\begin{eqnarray}
h=\begin{pmatrix}
1&2&-i\\
2&2&i\\
i&-i&1
\end{pmatrix}
\end{eqnarray}
We perform non-selective measurements as described earlier, at times $t=1,2,...$.
In this case the time evolution matrices between successive measurements are
\begin{eqnarray}
V_{12}=V_{23}=...=\exp (ith);\;\;\;t=1.
\end{eqnarray}
From this follows that
\begin{eqnarray}
{\mathfrak D}_{12}={\mathfrak D}_{23}=...={\mathfrak D}=
\begin{pmatrix}
0.232&0.223&0.545\\
0.223&0.551&0.226\\
0.545&0.226&0.229
\end{pmatrix}.
\end{eqnarray}
The eigenvalues of this matrix are $e_0=1$, $e_1=0.325$, $e_2=-0.315$.

If the probability vector at $t=1$ is ${\bf x}^{(1)}$, then at $t=s$ the probability vector is 
\begin{eqnarray}
{\bf x}^{(s+1)}={\bf x}^{(1)}{\mathfrak D}^s.
\end{eqnarray}
As $s\rightarrow\;\infty$ the ${\mathfrak D}^s$ goes close to ${\mathfrak U}$, and the ${\bf x}^{(s+1)}$ close to ${\bf u}$.
In this example, for $s>5$ the ${\mathfrak D}^s\approx {\mathfrak U}$ and the ${\bf x}^{(s+1)}\approx {\bf u}$.
Therefore the system is ergodic, and for large $s$ it follows all discrete paths with the same probability $1/3^s$.

\subsection{Rapidly repeated non-selective measurements: quantum Zeno effect and freezing of the paths}
We consider the homogeneous case with rapidly repeated non-selective measurements. Then ${\mathfrak D}$ is infinitesimally close to ${\bf 1}$, and it can be written as ${\mathfrak D}={\bf 1}+{\mathfrak E}$ where
\begin{eqnarray}
&&{\mathfrak E}=\begin{pmatrix}
-\epsilon(0,0)&\epsilon (0,1)&\cdots &\epsilon(0,d-1)\\
\epsilon (1,0)&-\epsilon(1,1)&\cdots &\epsilon(1,d-1)\\
\vdots&\vdots&\vdots&\vdots\\
\epsilon(d-1,0)&\epsilon(d-1,1)&\cdots&-\epsilon(d-1,d-1)\\
\end{pmatrix}\\
&&\sum _{b\ne a}\epsilon(a,b)-\epsilon(a,a)=\sum _{a\ne b}\epsilon(a,b)-\epsilon(b,b)=0.
\end{eqnarray}
Here $\epsilon(a,b)$ are infinitesimal non-negative numbers.
For example, we assume that the time evolution matrix for one of the intervals, is
\begin{eqnarray}
V=\exp(ith)\approx {\bf 1}+iht
\end{eqnarray}
 where $h$ is the Hamiltonian (with matrix elements $h(a,b)$) and $t$ is an infinitesimal time interval.
 Then
\begin{eqnarray}
\epsilon (a,a)=1-[h(a,a)t]^2;\;\;\;\epsilon (a,b)=|h(a,b)t|^2;\;\;\;a\ne b.
\end{eqnarray}

Then ${\mathfrak D}^s\approx {\bf 1}+s{\mathfrak E}$ and 
\begin{eqnarray}
{x}_a^{(s+1)}={x}_a^{(1)}[1-s\epsilon (a,a)]+s\sum _{b\ne a}{x}_b^{(1)}\epsilon (b,a)
\end{eqnarray}
If  $\epsilon _{\rm max}$ is the maximum of all $\epsilon (a,b)$, then for large $s$ which obeys the inequality
\begin{eqnarray}\label{tt}
1\ll s\ll \frac{1}{\epsilon _{\rm max}}
\end{eqnarray}
the probability vector ${\bf x}^{(s+1)}$ is almost equal to ${\bf x}^{(1)}$:
\begin{eqnarray}
{\bf x}^{(s+1)}\approx {\bf x}^{(1)}.
\end{eqnarray}
Consequently the corresponding universal convex polytope remains the same.

We also calculated the probabilities for discrete paths given in Eq.(\ref{path}).
We find that the probability that $\wp(1,s+1)$ is the `frozen path' $(a,...,a)$ is 
\begin{eqnarray}
q(a,...,a)=x_a^{(1)}[{\mathfrak D}(a,a)]^s=x_a^{(1)}[1-s\epsilon(a,a)]\approx x_a^{(1)}
\end{eqnarray}
The probability for any path that involves at least two different states is almost zero, because the ${\mathfrak D}(a,b)$ with $a\ne b$ is infinitesimal.
For example, for the path where the first $k$ states are $a$, and the rest of the states are $b$, the probability is  
\begin{eqnarray}
q(a,...,a,b,...b)&=&x_a^{(1)}[{\mathfrak D}(a,a)]^{k-1}{\mathfrak D}(a,b)[{\mathfrak D}(b,b)]^{s-k}\nonumber\\&=&
x_a^{(1)}[1-(k-1)\epsilon(a,a)]\epsilon(a,b)[1-(s-k)\epsilon(b,b)]\nonumber\\&\approx &x_a^{(1)}\epsilon(a,b)\approx 0
\end{eqnarray}
If there are more than two different states in the path, then the corresponding probability involves product of many infinitesimals and is even more close to zero.
For large $s$ (which obeys the inequality (\ref{tt})) the discrete path entropy (in Eq.(\ref{36})) goes to zero:
\begin{eqnarray}
E_{\wp}(1,s+1)=-\frac{1}{s+1}\sum _aq(a,...,a) \log q(a,...,a)=-\frac{1}{s+1}\sum _a x_a^{(1)}\log x_a^{(1)}\rightarrow 0.
\end{eqnarray}
It is seen that for time intervals that satisfy the inequality in Eq.(\ref{tt}), rapidly repeated non-selective measurements `freeze' (stroboscopically) the paths of the system.
This is the well known quantum Zeno effect \cite{Z1,Z2,Z3,Z4}, presented here in the language of Markov chains with doubly stochastic matrices, that has been used in this paper.
The fact that the path entropy goes to zero, is also an indication of the `freezing' of the paths.

\section{Discussion}

We studied Markov chains with doubly stochastic transition matrices.
Entropic quantities have been used to describe the randomness of the probability vectors and of the discrete paths. 
We have also introduced the universal convex polytopes in Eq.(\ref{25}) which contain all future probability vectors.
We have shown that as the step number (discrete time) increases the convex polytopes shrink (Eq.(\ref{40})),
and the minimum entropy of the probability vectors in them increases.
Ergodic Markov chains have been studied in section \ref{sec89}.

Quantum Mechanics is not a Markovian theory.
This is due to the non-diagonal elements in the density matrix, which are related to the superposition principle. 
We have discussed this in detail in sections \ref{sec78} and \ref{sec79}.
However non-selective measurements destroy the non-diagonal elements.
We have considered a quantum system which is forced stroboscopically by non-selective measurements, to be in a classical probabilistic mixture of states.
Between the measurements the system evolves unitarily in a quantum mechanical way.
We have shown that this system is described by time-dependent finite-state Markov chains with  doubly stochastic transition matrices.
We have explained that the quantumness of the system between successive measurements, makes the transition matrices doubly stochastic.
We note again here that generalisations of the classical Markov models called quantum Markov models, which are suitable for open quantum systems have been studied in \cite{AC1,AC2,AC3,AC4,AC5}.

We applied our results for Markov chains with doubly stochastic transition matrices, to our quantum system with the non-selective measurements.
In this quantum context, we have also introduced universal convex polytopes related to the diagonal density matrices after the measurements.
They contain all future probability vectors related to the measurements.
We have shown that as the discrete time increases the convex polytopes shrink (Eq.(\ref{d2})),
and the minimum entropy of the probability vectors in them increases.

Various examples have been discussed in section \ref{secww}.
We have considered a system which is initially in a position state, in a momentum state, and in the state $\rho=\frac{1}{d}{\bf 1}$.
In the ergodic case the system follows asymptotically all discrete paths with the same probability.
For rapidly repeated non-selective measurements we have shown how our approach which is based on Markov chains with doubly stochastic matrices,
leads to  the well known quantum Zeno effect.

Nonselective measurements on a quantum system, have recently been used in the literature for various objectives
(quantum communications\cite{N,N0}, quantum control\cite{N1}, quantum Zeno effect \cite{Z1,Z2,Z3,Z4}, quantum thermodynamics\cite{N2,N3},  etc).
In the present paper a sequence of non-selective measurements  has been studied as a `classical-quantum system', which immediately after the measurements is 
a classical probabilistic system,  and in the intervals between successive measurements is a quantum system that evolves with a unitary transformation.
We have shown that it is a rich system that exhibits various phenomena in various limits.

\newpage

\end{document}